# Incorporation of the Ce$^{3+}$ activator ions in LaAlO$_3$ crystals: EPR and NMR study

V. Laguta[1,3], M. Buryi[1], J. Pejchal[1], K. Uličná[2], V. Římal[2], V. Chlan[2], H. Štěpánková[2], Yu. Zagorodniy[3], M. Nikl[1]

[1]*Institute of Physics of the Czech Academy of Sciences, Cukrovarnicka 10, 16200 Prague, Czech Republic*

[2] *Faculty of Mathematics and Physics, Charles University, V Holešovičkách 2, 180 00 Prague 8, Czech Republic*

*Institute for Problems of Materials Sciences, National Academy of Sciences of Ukraine, Krzhizhanovsky 3, 03142 Kyiv, Ukraine*

**Abstract**

This work reports the results of EPR and NMR study of the Ce$^{3+}$ incorporation in LaAlO$_3$ single crystals grown by the micro-pulling-down method in the range of the Ce concentrations from x=0.1 at.% up to 100 at.%. From EPR measurements, Ce$^{3+}$ **g** tensor parameters were obtained as function of Ce concentration. The **g** tensor has orthorhombic symmetry even in the trigonal phase (x < 10 at.%) suggesting that the incorporation of Ce at La site lowers lattice symmetry near this ion. The local properties of the La$_{1-x}$Ce$_x$AlO$_3$ crystals were further studied by $^{27}$Al and $^{139}$La high-resolution NMR measurements. It was found that $^{139}$La chemical shift has the Fermi contact interaction origin. It linearly increases with Ce concentration up to 165 ppm at x = 0.5. Due to this strong Fermi contact interaction, separated peaks corresponding to different Ce-O-La spin transfer passways are resolved in the $^{139}$La NMR spectra. On the other hand, no Fermi contact interaction is visible in $^{27}$Al NMR spectra. However, these spectra contain satellite peak which intensity linearly increases with increase of Ce concentration leaving position of this peak unchanged. This was interpreted as manifestation of crystal structure modification in vicinity of Ce ions in agreement with EPR data. Thus, optical properties of Ce$^{3+}$ in LaAlO$_3$ will be namely determined by the local crystal structure near this ion.

**1. Introduction**

The search of new materials with heavy ions suitable for applications as scintillators is an important topic of today physics and chemistry. Presently scintillators have been employed in high-energy physics, environmental monitoring, industrial defectoscopy, geological survey and oil well logging, astronomy and others [1,2]. One of the promising class of such scintillation materials belongs to aluminium perovskites [3]. The most known example of the perovskite scintillators is the Ce-doped YAlO$_3$. However, it has low density and effective atomic number of 5.37 g/cm$^3$ and 33.5, respectively, and the resulting low stopping power for high-energy photons is the major disadvantage of this material. In this respect, LuAlO$_3$ (LuAP) perovskite with much higher density 8.4 g/cm$^3$ and effective atomic number Z$_{eff}$=64.9 could be highly promising material. However, due to a severe instability of the perovskite phase in the process of crystal growth from the melt, resulting in a high price for the crystals, mixed $(Lu_{0.7}Y_{0.3})$AP:Ce scintillators were chosen for industrial-scale production [4].

Another perovskite material with also high density is the compositional analogue $LaAlO_3$ (LaAP), whose crystal growth is much easier due to almost ideal value of Goldschmitt tolerance factor [5]. Its growth technology [6] is well established as this crystal extensively uses as substrate for thin films epitaxial growth [7]. There are a number of papers which devoted optical absorption and luminescence of Ce-doped $LaAlO_3$, e.g. [6,8,9,10]. While some crystals with low Ce concentration (up to 1 at.%) show characteristic $Ce^{3+}$ 4f-5d transition emission [6,8], in general, the radioluminescence efficiency of the Ce-doped LaAP crystals was found to be very low. This was explained as due to strong ionization of the $Ce^{3+}$ 5d excited states which are all located in the conduction band [11].

The most studied rare-earth impurity in $LaAlO_3$ is $Gd^{3+}$ which Electron paramagnetic resonance (EPR) spectra were measured and analyzed in details in both single crystals [12] and powders [13]. This ion substitutes for La at simple trigonal symmetry. $LaAlO_3$ and $GdAlO_3$ create $Gd_{1-x}La_xAlO_3$ solid solution where the band gap gradually decreases with increase of La content [14]. $LaAlO_3$ also creates similar solid solution with $CeAlO_3$ [15]. However, to the best of our knowledge, $Ce^{3+}$ EPR spectrum was never measured in $LaAlO_3$ crystals or even in the $La_{1-x}Ce_xAlO_3$ solid solution in spite that in other aluminum perovskites (e.g. $YAlO_3$, $LuAlO_3$) the $Ce^{3+}$ EPR is well visible [16,17]. Because the optical properties of Ce-doped $LaAlO_3$ are still not elucidated we decided to perform detailed investigations of $Ce^{3+}$ EPR in the $La_{1-x}Ce_xAlO_3$ solid solutions in the range of Ce concentration from x=0.001 up to x=1.0. Incorporation of Ce ions into $LaAlO_3$ lattice was also studied via measurements of $^{139}La$ and $^{27}Al$ solid-state Nuclear magnetic resonance (NMR). NMR method is of particular useful in quantification of the site occupancy of atoms in a material and in study of chemical bonds. In particular, NMR showed that the $Ce^{3+}$ spin transfer (and thus charge transfer) to $La^{3+}$ is much stronger than to $Al^{3+}$. Both EPR and NMR also suggest $LaAlO_3$ crystal structure transformation in vicinity of Ce ions even at low their concentration in the trigonal phase.

## 2. Experimental methods

$(LaAlO_3)_{1-x}(CeAlO_3)_x$ (x=0.001; 0.005; 0.01; 0.05; 0.1; 0.2; 0.3; 0.5; 1.0 and defined as LaAP:xCe) crystals were grown by a micro-pulling-down method in the reducing $Ar+5\%H_2$ atmosphere. The details of the growth procedure is described in [8]. Some of these crystals were previously exanimated by X-ray diffraction, micro-X-Ray fluorescence, optical absorption, luminescence and radioluminescence measurements [8]. All these crystals have perovskite structure as confirmed by the X-ray diffraction. However, as suggested in [18], a small amount (below the X-ray diffraction sensitivity) of other aluminate phases is not excluded. Their presence was confirmed by scanning electron microscopy for some of samples studied in [18].

EPR measurements were performed in the X-band (9.4 GHz) with a commercial Bruker EMX spectrometer at the temperatures 6-296 K. $^{27}Al$ static and magic-angle spinning (MAS) NMR spectra and $^{139}$La MAS NMR spectra of powdered samples were recorded at room temperature in the magnetic field of 11.75 T using a wide-bore Bruker Avance III HD spectrometer ($^{27}$Al Larmor frequency ≈ 130.4 MHz, $^{139}$La Larmor frequency ≈70.7 MHz), with a nonselective single-pulse excitation. The samples were packed into ZrO$_2$ rotors of 2.5 mm in diameter and were rotating at 20 kHz MAS speed.

LaAP at T > 813 K has simple cubic perovskite structure belonging to the *Pm-3m* space group [19]. Below T$_C$ = 813 K, it undergoes phase transition to the rhombohedral *R-3c* structure related to rotation of the oxygen octahedrons around one of the [111] trigonal axes of the cubic phase through the same angle $\phi$ [19], with half of the polyhedral rotating clockwise around the trigonal axis, and half rotating anticlockwise (Fig. 1). The rotation angle is quite large below 50 K, it reaches the value of 5.7$^0$.

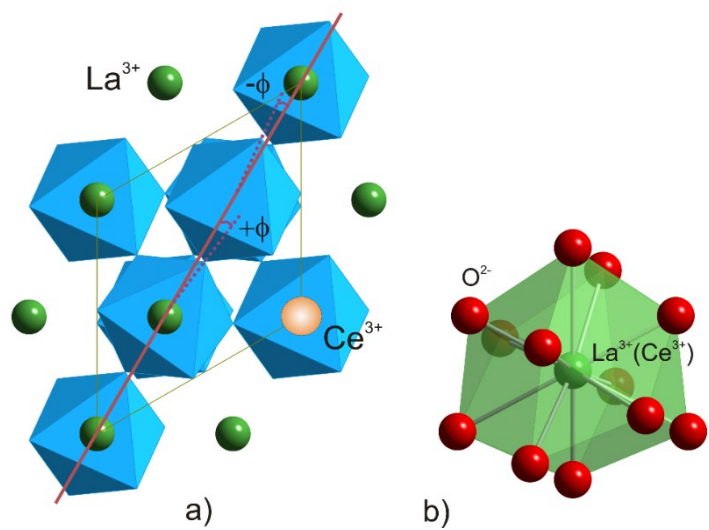

Fig.1. (a) LaAlO$_3$ crystal structure viewed down the *c* trigonal axis. The oxygen polyhedral rotate through the angle $\phi$ clockwise and anticlockwise around the trigonal axis. Ce$^{3+}$ substitutes for La$^{3+}$. (b) Local surroundings of La$^{3+}$(Ce$^{3+}$) ions. Oxygen ions create four distorted but equivalent triads.

There is only one crystallographic site (6*a* sites) for La cations. Ce$^{3+}$ substituting for La$^{3+}$ has respectively also one crystallographic position in the lattice as illustrated in Fig. 1. This ion has outer electronic shell 4$f^1$ and electron spin $S$ = 1/2.

In contrast to LaAlO$_3$, the sequence of the reversible phase transitions *I*4/*mcm* ↔ *Imma* ↔ *R-3c* ↔ *Pm-3m* has been detected in CeAlO$_3$ and solid solutions formed in the pseudo-binary system CeAlO$_3$–LaAlO$_3$ [15]. Below approximately 10 K, where the Ce$^{3+}$ EPR spectra can be measured, the lattice symmetry in the (1-x)LaAlO$_3$–xCeAlO$_3$ solid solutions changes from the trigonal *R-3c* to orthorhombic *Imma* and finally to the tetragonal *I*4/*mcm* as the CeAlO$_3$ contents increases to approximately x=0.5 [15].

### 3.1. Results and discussion
#### 3.1.1. Ce$^{3+}$ EPR spectra in LaAP:Ce

As an example, Fig. 2 presents EPR spectra measured in selected LaAP:Ce crystals at different temperatures. At low Ce concentration (0.001Ce and 0.005Ce; 0.1% and 0.5%, respectively) the EPR spectra mainly contain narrow $Gd^{3+}$ resonances, which spectral parameters are well studied in LaAP [12,13]. The expected for these Ce concentrations very strong and narrow $Ce^{3+}$ spectral lines are not visible. Only in the crystal with the Ce concentration 0.5% two characteristic but very broad lines appear below 9 K (Fig. 2), which could belong to $Ce^{3+}$, in spite that $Gd^{3+}$ lines are narrow, indicating good crystalline quality of the crystal. The $Ce^{3+}$ spectral lines are thus expected to be very narrow like, for instance, in $YAlO_3$ [16], where the $Ce^{3+}$ linewidth is only 13-20 G for the same Ce concentration. This fact speaks in favor that $Ce^{3+}$ resonance lines are essentially broadened by the magnetic dipole interaction between spins located close to one other but are not broadened by crystal imperfections and, thus, the Ce ions do not homogeneously distributed in crystal but prefer to create Ce-rich regions. Further increase of the Ce concentration leads to substantial increase of the $Ce^{3+}$ EPR intensity without marked change in linewidth (LaAP:0.05Ce sample, Fig. 2).

A drastic change in the $Ce^{3+}$ spectrum is seen at the Ce concentration of 20% (LaAP:0.2Ce sample, Fig. 2). The spectrum is split into many components which are sited on a very broad line at 3-7 kG magnetic field region. This corresponds to change of the crystal structure of the $La_{1-x}Ce_xAlO_3$ solid solution from the rhombohedral *R-3c* to the tetragonal *Imma* phase at x>0.1 [15]. Numerous EPR transitions from $Ce^{3+}$ - $Ce^{3+}$ dimers appear as well that makes the spectrum analysis practically impossibly.

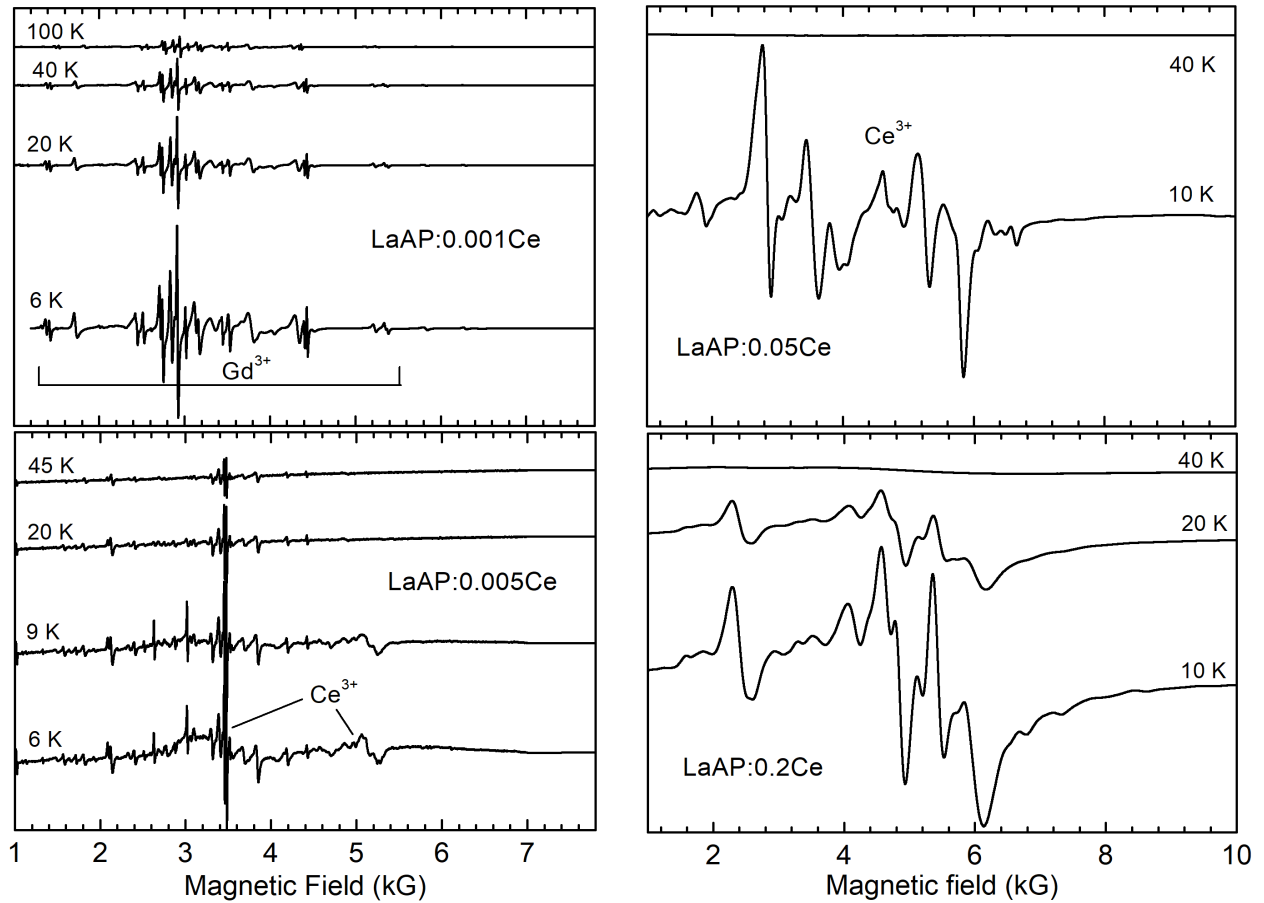

Fig. 2. EPR spectra measured in the LaAP:Ce crystals with selected Ce concentrations at different

temperatures and crystals orientation close to the trigonal axis. The broad $Ce^{3+}$ spectral lines are indicated. At low Ce concentration (x=0.001Ce) only narrow $Gd^{3+}$ resonances are seen.

Because our LaAP:Ce crystals were not singledomain, it was problematically correct determination of the $Ce^{3+}$ g factors from EPR spectra measured in crystals only. Therefore, EPR spectra were also measured in powders of grinded crystals. The EPR spectra measured in the grinded crystals for Ce concentration from x=0.005 up to x=0.5 are shown in Fig. 3a. In powders, EPR spectrum is averaged over all random orientations of crystallites. Only characteristic features (peaks) are presented in spectra at the fields corresponding to g-factors principal values [20] as it is well seen in the spectrum for LaAP:0.05Ce sample and its simulation (Fig. 3a). The simulation was performed by using the computer program "Powder" [21] and the spin Hamiltonian for a paramagnetic ion with the spin $S = 1/2$:

$$H_1 = \mu_B \mathbf{BgS}, \tag{1}$$

where **B** is magnetic field, **g** is g tensor with three principal values $g_1$, $g_2$, and $g_3$, **S** is the spin operator with three components along g tensor principal axes. The program "powder" uses bispiral approach to obtain directions homogeneously distributes in sphere. For the search of resonance magnetic fields and transition probabilities for reference directions, exact diagonalization of spin Hamiltonian matrices is used [21].

It is reasonable to assume that one of the principal axis (corresponding to $g_3$ value) is oriented along the trigonal axis for the samples with low Ce concentration (*R-3c* symmetry) in analogy with $Ce^{3+}$ EPR in $YAlO_3$ crystal [22], which crystal structure is very close to that of $LaAlO_3$. It must rotate to the [001] tetragonal axis for samples with high Ce concentration according to the *I4/mcm* symmetry at x ≈ 0.5. But the explicit orientations of the principal axes could not be determined from our data due to multidomain structure of our crystals. Determined g factors values for all studied compositions are listed in Table I.

Table I. $Ce^{3+}$ g factor values in LaAP:xCe determined from $Ce^{3+}$ EPR powdered spectra at 6-10 K.

| xCe | $g_1$ | $g_2$ | $g_3$ | Lattice symmetry |
|---|---|---|---|---|
| 0.01 – 0.05 | 0.960(2) | 1.935(5) | 2.340(2) | *R-3c* |
| 0.1 | 0.97(1) | 1,87(1) | 2.43(1) | *R-3c* |
| 0.2 | 1.06(1) | 1.37(1) | 3.02(1) | *Imma* |
| 0.3 | 1.04(1) | 1.32(5) | 3.11(1) | *Imma* |
| 0.5 | 1.02(2) | 1.02(2) | 3.04(2) | *I4/mcm* |
| 1.0 | 1.71(2) | 1.71(2) | 2.56(2) | *I4/mcm* |

One can see that the $Ce^{3+}$ spectrum even in the trigonal phase (x = 0.01 –0.1) is described by three non-equal g factor components that corresponds to the orthorhombic local symmetry at Ce ion, i.e.

incorporation of Ce at La site lowers lattice symmetry in vicinity of the Ce ion. This is in contrast with Gd$^{3+}$ ions ($4f^7$, S=7/2) which also substitute for La but their EPR spectra are described assuming the pure trigonal (axial) symmetry [12]. The same takes place for Nd$^{3+}$ ions ($4f^3$, S=1/2) which EPR spectrum is also axial ($g_\parallel$ = 2.12, $g_\perp$ =2.68) [23]. At higher Ce concentration (0.5 > x > 0.1), symmetry of the lattice becomes orthorhombic *Imma* and g tensor corresponds to this symmetry as well as in the tetragonal *I4/mcm* phase at x > 0.5, i.e., Ce$^{3+}$ EPR spectrum is described by three non-equal components.

Note that a narrow peak at 3.2 kG seen in LaAP:0.005Ce and LaAP:0.01Ce samples originates from unidentified defect. It is visible only at T<9 K (Fig. 3b, upper panel) and is not present at Ce concentration larger about 5% (Fig. 3b, down panel), when the Ce$^{3+}$ spectrum becomes dominated in intensity. Its g factor is equal to 2.115 that is very close to g factor value measured for Nd$^{3+}$ in LaAP [23]. Therefore, most probably, this background impurity is Nd$^{3+}$.

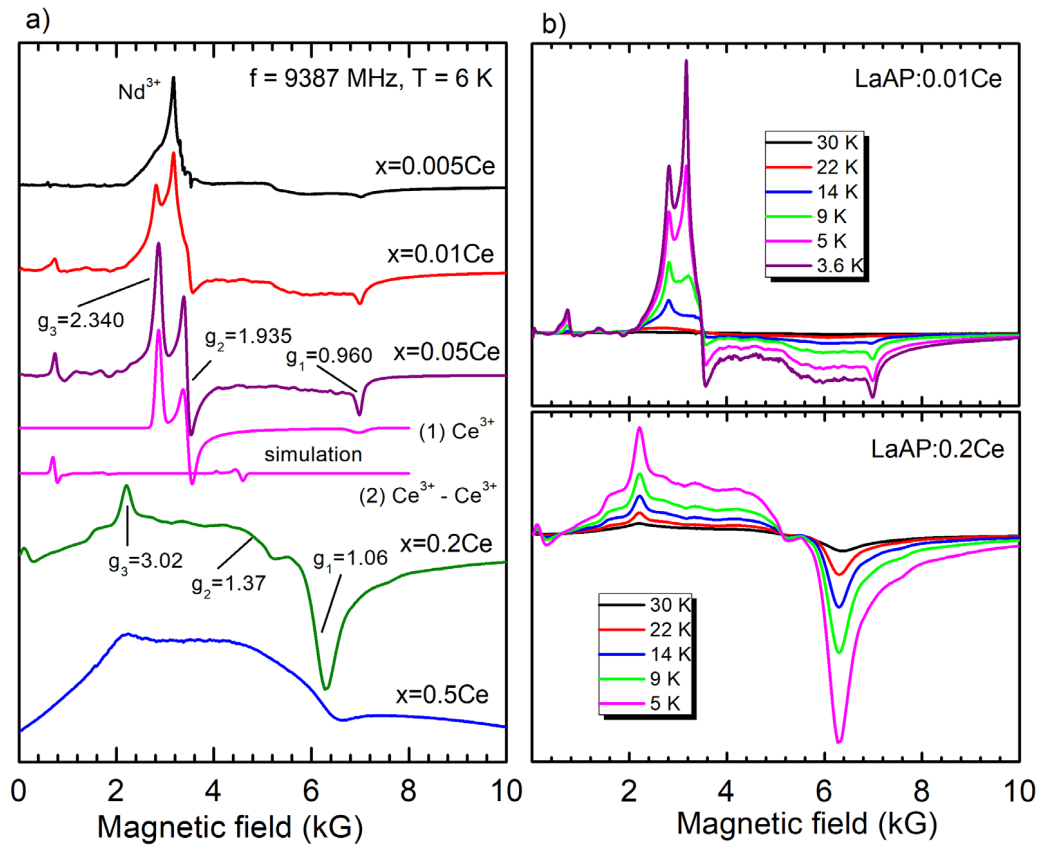

Fig. 3. (a) Ce$^{3+}$ EPR spectrum in LaAP:xCe grinded crystals as a function of the Ce concentration x. The simulated spectrum is shown for the LaAP:0.05Ce sample: spectra (1) and (2) correspond to isolated Ce$^{3+}$ and Ce$^{3+}$ - Ce$^{3+}$ exchange coupled pairs, respectively. (b) Temperature dependence of the Ce$^{3+}$ spectrum in LaAP:0.01Ce and LaAP:0.2Ce samples. A sharp lines in the LaAP:0.01Ce sample at 3.2 kG is created most probably by Nd$^{3+}$ impurity.

However, the peak at low-field side (75-700 G, depending on Ce concentration) surely belongs to Ce$^{3+}$ ions. It appears at high Ce concentration and shifts similar as the main Ce peaks with increase of Ce concentration. It is reasonable to describe this peak as well as other low-intensity peaks to the Ce$^{3+}$

- $Ce^{3+}$ exchange coupled pairs. To prove this assumption, we calculated powder spectrum expected from the $Ce^{3+}$ - $Ce^{3+}$ dimers. The following spin Hamiltonian is used:

$$H = \mu_B \mathbf{S_1 g_1 B} + \mu_B \mathbf{S_2 g_2 B} + J_z S_{1z} S_{sz} + J_x S_{1x} S_{2x} + J_y S_{1y} S_{2y}, \tag{2}$$

where $J_i$ are the spin-spin interaction constants including the contribution from both the dipole-dipole and anisotropic exchange interactions. Because only one crystallographic site exists for Ce ions, we can take $\mathbf{g_1} = \mathbf{g_2} = \mathbf{g}$. The spin Hamiltonian (2) assumes that both $\mathbf{g}$ and spin-spin interaction tensors have the same principal axes, which could not be completely valid as the explicit orientation of the $\mathbf{g}$ tensor axes is unknown. Therefore, the spin-spin interaction constants can be considered as effective quantities renormalized to the common with g tensors axes. However, because g tensors are fixed in the powder spectrum simulation, the overall error in the $J_i$ values is expected to be in the range of 20-25%.

The simulation of the $Ce^{3+}$ - $Ce^{3+}$ pairs in the LaAP:0.05Ce sample is shown in Fig. 3a, spectrum (2). The calculated spectrum well describes the main peak at 700 G and few other weak peaks at 1900 and 4050 G. The following spin-spin interactions constants were determined: $J_z = -1770 \times 10^{-4}$ cm$^{-1}$, $J_x = 1490 \times 10^{-4}$ cm$^{-1}$, $J_y = 280 \times 10^{-4}$ cm$^{-1}$. They are in the range measured for $Ce^{3+}$ pair centers in YAlO$_3$ [16] and in LaCl$_3$ [24], for instance, ($|J_z| = 0.1 - 0.6$ cm$^{-1}$) taking into account that in both these crystals the largest g factor is almost two times bigger than that in LaAlO$_3$ (dipole interaction is proportional to $g^2$).

As the Ce concentration increases to x=0.3–0.5, contribution of the $Ce^{3+}$ pairs as well as $Ce^{3+}$ triads and complexes with more number of Ce ions coupled by the dipole and exchange interaction becomes dominated over the contribution of single Ce ions into spectrum. As a result, the $Ce^{3+}$ spectral lines become very broad. Moreover, the anisotropy in the spectrum decreases due to exchange averaging [25] that is reflected in the decrease of the $g_3 - g_1$ difference (Table I).

### 3.1.2. $^{139}$La NMR study

LaAlO$_3$ – CeAlO$_3$ solid solutions contain $^{27}$Al (nuclear spin $I = 5/2$, magnetic dipolar moment $\mu$ expressed in nuclear magnetons $\mu = 4.309$ $\mu_N$, electric quadrupolar moment $Q = 1.466$ barn) and $^{139}$La ($I = 7/2$, $\mu = 3.156$ $\mu_N$, $Q = 0.200$ barn) nuclei suitable for NMR measurements. Both these nuclei have natural abundance of 100% and large magnetic moments. Their NMR spectra are thus intensive and informative. NMR method is often used to quantify the site occupancy of atoms in a material and in study of chemical bonds [26]. NMR frequency is also very sensitive to hyperfine fields produced by paramagnetic ions [27,28]. Therefore, NMR is very desirable in the investigation of Ce ions incorporation into LaAlO$_3$ lattice.

$^{139}$La MAS NMR spectra of the central $1/2 \leftrightarrow -1/2$ transition measured at room temperature with 20 kHz rotation are shown in Fig. 4a for crystals grinded into powder with the Ce content x = 0.005, 0.01, 0.2, 0.5 and 1. The spectra of x = 0.005 and 0.01Ce samples have a similar shape. The shape corresponds to a line broadening induced by electric quadrupolar interaction of $^{139}$La quadrupole

moment with electric field gradients (EFG) as the second order perturbation of Larmor frequency. Separately, $^{139}$La MAS NMR spectrum of x=0.005Ce sample is shown in Fig. 4b together with the fitted spectrum performed in the Bruker "SOLA" program [29]. From the fitting, the quadrupole constant $C_q = \dfrac{e^2 V_{zz} Q}{h} = 6.95\,\text{MHz}$ and asymmetry parameter $\eta = (V_{xx} - V_{yy})/V_{zz} \approx 0$ were obtained. Quadrupole parameters are in agreement with those measured in single crystal using quadrupole satellite transitions [30]. The fit giving these EFG parameters seem fairly plausible, thought it is evident that there is also another very weak contribution on the left side of the central line which can be assigned to the presence of Ce. This feature increases with the Ce content increase.

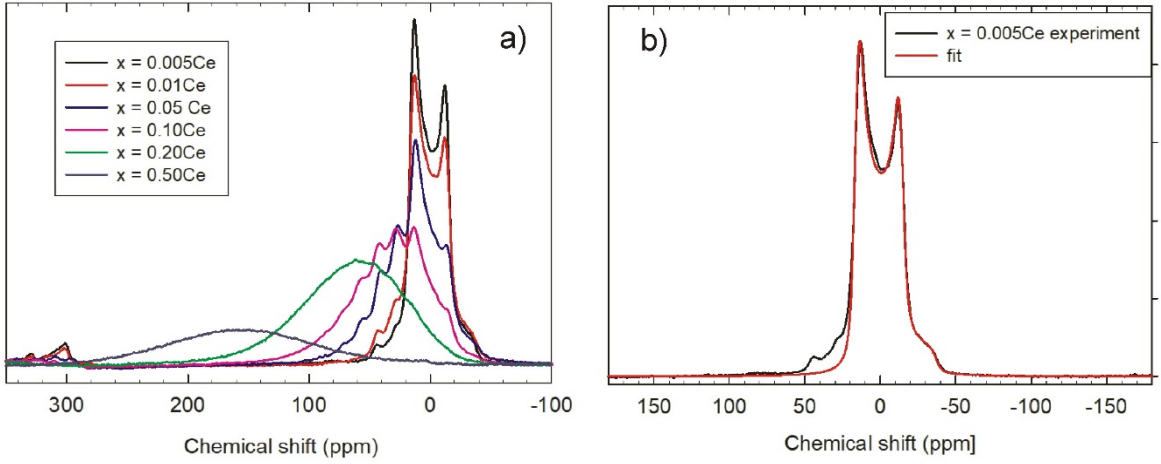

Fig. 4. (a) $^{139}$La MAS NMR spectra measured in the LaAP:xCe samples; (b) Experiment (black line) and one-site fit (red line) of $^{139}$La MAS central transition line for x = 0.005Ce.

It is evident from the measurements of the central transition in samples with Ce up to x = 0.1 that the spectrum is composed of several contributions, with similar or very close EFG parameters and nearly equidistant frequency shifts. Their relative integral intensities naturally depend on Ce content. The particular contributions should correspond to the number of substituted $Ce^{3+}$ ions in the neighborhood of resonating $^{139}$La nuclei.

The central lines in the spectra of x=0.2 and x=0.5Ce samples are markedly broadened (linewidth ~100 ppm for 20 % which is approximately 2.5 times broader than for low amount of Ce, and ~200 ppm for x=0.5Ce) and quite strongly shifted (by ~60 ppm and ~170 ppm, respectively) downfield. The lines lack any resolved spectral features. The shift can be assigned to the change of isotropic part of chemical shift induced by the presence of paramagnetic $Ce^{3+}$ in the neighborhood of resonating La. Indeed, the positive frequency shift has paramagnetic origin (Fermi contact interaction) and is proportional to concentration of Ce paramagnetic ions. It can be usually written as [27]:

$$\delta_0 = -\dfrac{A}{f_0}\langle S_z \rangle, \tag{3}$$

where $f_0$ is the Larmor frequency, $A$ is electron-nuclear interaction constants of the nuclear spin with electron spin (it is proportional to the electron spin density at the nucleus), $\langle S_z \rangle$ is the time-averaged electron spin projection along the quantization axis. Because the Fermi contact interaction is generally additive, the overall paramagnetic shift can be obtained by summing the shifts induced by each magnetic ion, so it should be proportional to concentration of paramagnetic ions. Indeed, the shift of the center-of-gravity of $^{139}$La spectrum linearly increases with Ce concentration (Fig. 5).

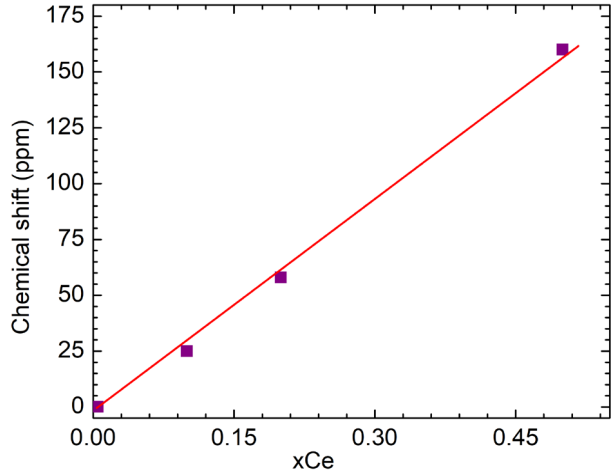

Fig. 5. Dependence of the $^{139}$La chemical shift (center-of-gravity of MAS spectral line) on Ce concentration.

To get more inside into chemical bonds in LaAP:xCe we performed simulation of the $^{139}$La spectra for x = 0.05 and 0.1Ce. For these Ce concentrations, well-resolved NMR peaks for different Ce-O-La bonds are well seen as the spectra are not too much broadened by Ce-La dipole interactions. Quadrupole parameters determined above were used in the simulation. In LaAP structure, the spin transfer to the closest La ions in each cube face (for instance, La1, La2, La3 in Fig. 6) can be realized via $180^0$ Ce-O-La bonds (as Ce-O1-La1 in Fig. 6) and $90^0$ Ce-O-La bonds (as Ce-O2-La2 type bonds in Fig. 6) with participation of one and four oxygens, respectively. Each La/Ce ion has 6 nearest La/Ce neighbors along six [100] pseudocubic directions and 12 next-nearest La/Ce neighbors along [110] pseudocubic directions.

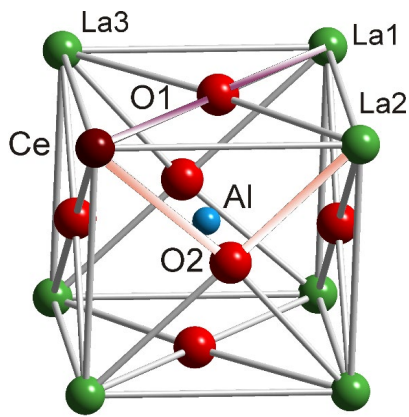

Fig. 6. Schematic depiction of LaAlO$_3$ lattice with Ce at La site. As an example, $180^0$ (Ce-O1-La1) and $90^0$ (Ce-O2-La2) bonds are indicated.

The contribution of different $^{139}$La nuclei into spectrum intensity was calculated with using the following relation:

$$P(n,k,x) = (1-x)^{6-n} x^n \left[(1-x)^{12-k} x^k\right] \frac{12!}{(12-k)!k!} \frac{6!}{(6-n)!n!}, \qquad (4)$$

where the distribution of Ce ions over 6 nearest and 12 next-nearest to La positions, the first and second coordination spheres, respectively, was assumed. Here $x$ is the Ce concentration, $n$ is number of Ce ions in the first coordination sphere, and $k$ is number of Ce ions in the second coordination sphere.

Usually, $90^0$ bonds give chemical shift about two times smaller as compared to $180^0$ bonds (at least, such relation exists for $^{27}$Al [26]). Therefore, twelve $180^0$ bonds gives the same chemical shift as twenty four $90^0$ bonds. This is also confirmed by the fact that peaks in spectra are shifted on the same value multiple of 14.35 ppm. Taking into account this fact, the chemical shift for each $P(n,k,x)$ probability value was calculated as $\delta(n,k) = (k+2n) \times 14.35$ ppm. Result of the calculation is shown in Fig. 7 for the Ce concentration $x = 0.05$ and 0.1. Satisfactory agreement between the measured and calculated spectra supports the proposed model of chemical bonds.

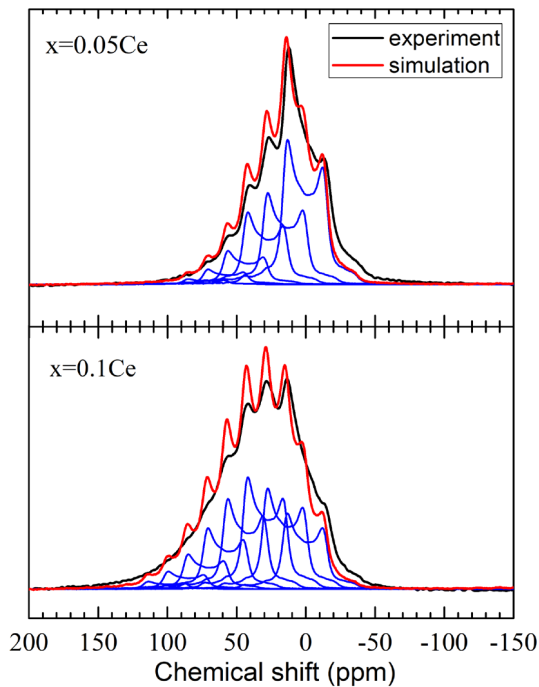

Fig. 7. Experimental (dark solid lines) and simulated (red solid lines) $^{139}$La MAS NMR spectra for Ce concentration 5 and 10%. Blue solid lines show contributions from different arrangements of Ce ions around resonating $^{27}$Al nucleus with probabilities according to Eq. (4).

Essential line broadening for the samples with 20 % and 50 % Ce can be qualitatively assigned to the distribution of chemical shift values and EFG parameters and connected to the distribution of configurations of La neighborhood with various number and arrangement of paramagnetic Ce$^{3+}$ ions, including also possible changes of lattice parameters. In addition, there is also anisotropic magnetic dipole interaction between nuclear and electron spins (also referred to as "pseudocontact" shift [31]), which could be only partly averaged by the magic-angle spinning. The dipole interaction also increases with increase of Ce concentration leading to broadening of NMR line.

### 3.1.3. $^{27}$Al NMR study

$^{27}$Al static NMR spectra for the lowest Ce contents x = 0.005 and 0.01 are shown in Fig. 8a. The spectra for these samples are typical for the nuclei with strong electric quadrupolar interaction, i.e. there, the central line and four resolved satellites of non-central transitions appear; all five lines in the spectrum are quite broad (the broadening of the central transition is mainly caused by anisotropic part of chemical shift and the second order quadrupole interaction). The maxima of the satellites lines for both samples are approximately at the same positions. The lines of x = 0.005Ce sample is higher than for x = 0.01Ce sample, which is consistent with the broader distribution of paramagnetic neighborhoods of Al when the content of Ce is higher. The fitting of the spectrum for x = 0.005Ce (Fig. 8b) gives $C_Q$ = 0.16 MHz and η = 0 in accordance with [30], where these parameters were directly obtained from single-domain crystal measurements. This, however, differs from $C_Q$ = 0.17 MHz and η = 0.6-0.8 given in [32]. The large asymmetry of EFG tensor η = 0.6-0.8 reported in [32] seems to be not reliable for LaAP as the EFG tensor must be axial at Al site (η= 0) due to the rhombohedral *R-3c* symmetry.

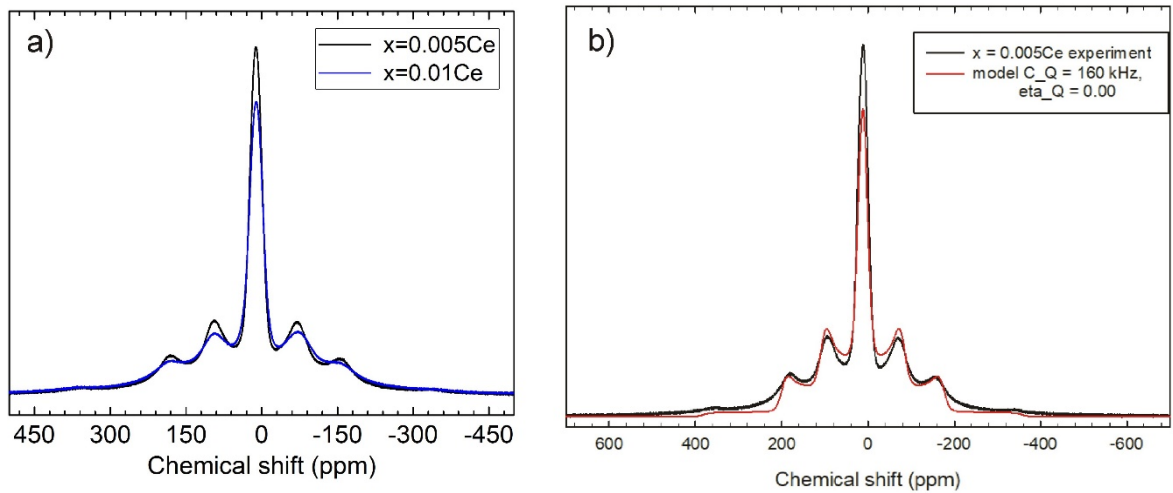

Fig. 8. (a) $^{27}$Al static NMR spectra measured in LaAP:xCe samples (x=0.005Ce and 0.01Ce) at room temperature. (b) Measured (black line) and calculated (red line) $^{27}$Al NMR spectrum for x = 0.005Ce sample.

For higher concentrations of Ce (x = 0.05 – 0.2), Fig. 9, position of the main maximum does not change visibly but the lines broaden due to increased dipole interaction with paramagnetic $Ce^{3+}$ ions. In addition to the line broadenig, an asymmetric feature (additional maximum) appears on the right side (upfield) from the central maximum. Its intensity grows with rising Ce content, but its position does not visibly change and it seems that no other additional maxima rise. This additional maximum (which makes the spectrum asymmetric) is shifted from the central maximum by about 90 ppm, that is 11.8 kHz.

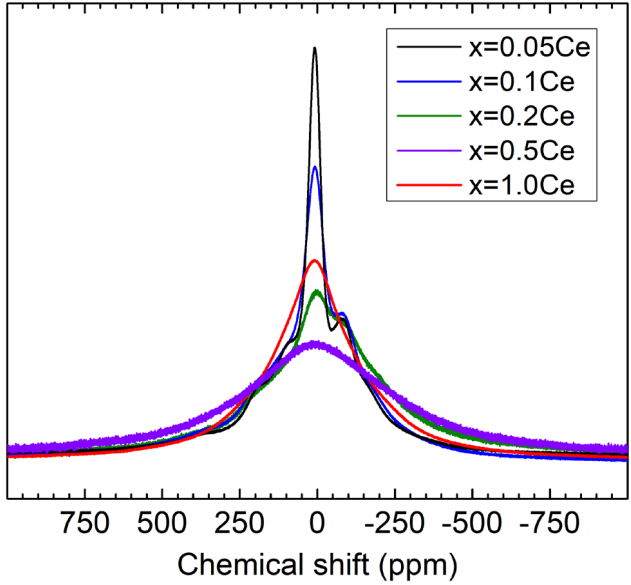

Fig. 9. $^{27}$Al static NMR spectra measured in LaAP:xCe samples (x=0.05 – 1.0).

This feature in $^{27}$Al spectrum can be explained as due to dipole interaction of Al nuclear spin with $Ce^{3+}$ electron spin. This electron-nuclear interaction is quite large for Al nuclei as the Ce-Al distance is short (3.272 and 3.281 Å) and the magnetic moment of $^{27}$Al nucleus is large. The influence of the dipole interaction on NMR spectrum can be considered as appearance of anisotropic chemical shift with the average amplitude:

$$\delta_{aniso} = (\mu_0 / 4\pi r^3)(2g_{iso}^2 \mu_B^2 S(S+1)/3kT), \qquad (5)$$

where $\mu_0 = 4\pi \cdot 10^{-7}$, $k = 1.38 \cdot 10^{-23}$ J/K, $r = 3.28 \cdot 10^{-10}$ m, $g_{iso}^2 = 4.04$. This gives for x = 0.2 and T = 300 K, $\delta_{aniso} \approx 120$ ppm, comparable in value with the experimentally measured shift of 90 ppm.

The spectra of the samples with the x=0.5 and 1.0 of Ce are broad, no spectral features are resolved. The spectrum for 50 % is broader than for 100%, which is consistent with the assumption that the atomic order in the 50% is poorer due to many different configurations of Al neighborhood. Also, the spectrum for x = 1.0 of Ce is partly exchange narrowed due to exchange interactions between $Ce^{3+}$ spins. This is well manifested in disappearance of broad wings at spectral line.

$^{27}$Al MAS NMR spectra for all Ce concentrations are shown in Fig. 10. The central-transition spectrum for the samples with 0.5 and 1.0 % of Ce consists of one "main" line from unperturbed Al nuclei at $\approx 11.8$ ppm (referenced to $^{27}$Al in $Al(NO_3)_3$) and a weak satellite at ~5.5 ppm. At first sight, the satellite could be assigned to the presence of Ce replacing one of the nearest La. The shift from the main line is about 5.5 ppm, which is comparable, e.g., with the $^{27}$Al NMR results obtained for $Y_3Al_5O_{12}$:Ce [33] where similar satellite line (assigned to Al perturbed by one Ce ion in the first coordination shell) is shifted from the main line at approximately 17 ppm up-field.

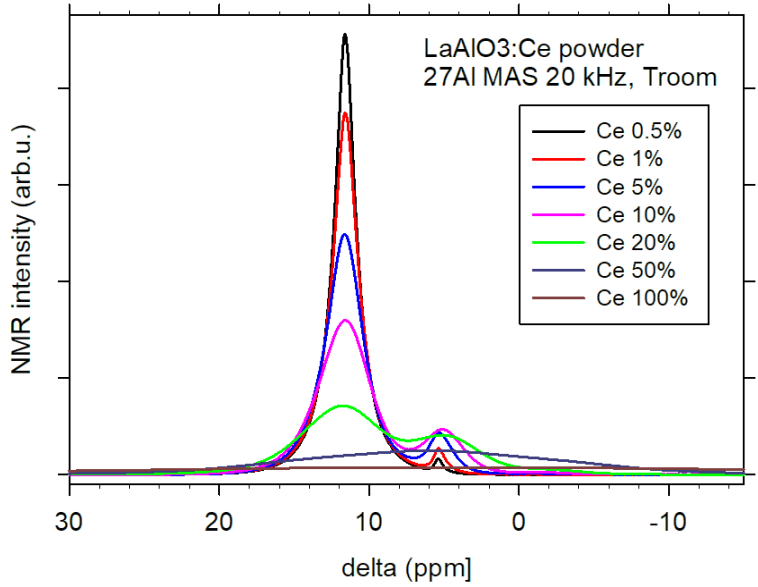

Fig. 10. $^{27}$Al MAS NMR spectra measured at room temperature and at rotation 20 kHz in LaAP:xCe samples.

However, the relative area of the satellite line with respect to the main line does not correspond to that expected for the random distribution of Ce ions. The relative integral intensity of the satellite line is simply proportional to Ce concentration (Fig. 11).

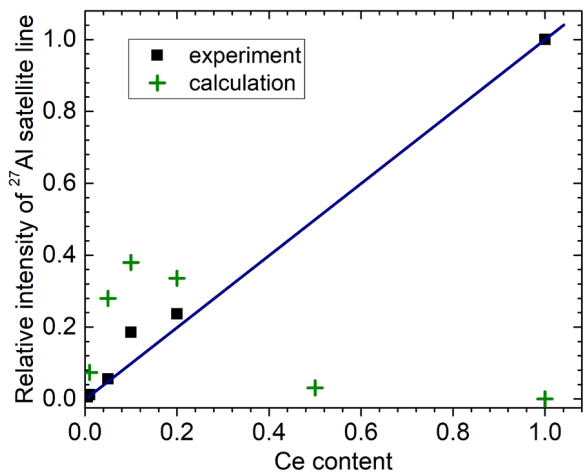

Fig. 11. Relative integral intensity of $^{27}$Al NMR satellite line at 5.5 ppm as a function of Ce content. Calculated data are presented for the case when only one Ce ions can be substituted for Al as the nearest neighbor.

At low Ce concentration (x ≤ 0.1), each Ce paramagnetic ion has 8 nearest Al neighbors. Therefore, if this satellite line has Fermi contact origin, its intensity $I$ must change with Ce concentration as $I/I_0 \approx 8x$, where $I_0$ is the intensity of the central unperturbed line. But it changes as $I/I_0 \approx x$. Besides, in contrast to $^{139}$La NMR, the frequency shift of the satellite line in the $^{27}$Al NMR is negative. Moreover, it does not markedly depend on Ce concentration. Therefore, this satellite line can not originate form Fermi contact interaction. This means that the spin transfer from $Ce^{3+}$ ions to $Al^{3+}$ ions has much lower efficiently as compared to spin transfer to $La^{3+}$ ions.

For x = (0.05 – 0.2)Ce also the second satellite appears at ≈ -0.2 ppm which is approximately at a twice as high distance from the main line, however again with much lower intensity than calculated from binomial distribution (supposing two Ce atoms replacing La in the neighborhood of Al and nominal Ce concentration). Its fraction for sample with x = 0.2 is only 6.6%. The spectra of the samples with 50% and 100% Ce content consist of one broad line, its maximum corresponds to the position of the first satellite in low-Ce concentration samples. The line for 100% Ce sample is the broadest one, in opposite to the static spectra. This confirms that the spectrum for 100% Ce sample is exchange narrowed. The broadening is thus partly homogeneous and line can not be much narrowed in MAS measurements.

The behavior of $^{27}Al$ spectrum in the $La_{1-x}Ce_xAlO_3$ solid solution thus suggests that the chemical shift for $^{27}Al$ nuclei has "structural" origin rather than magnetic origin, i.e. is related to change of crystal structure in the vicinity of Ce ions that was also reflected in the orthorhombic symmetry of the $Ce^{3+}$ **g** tensor. This structure (at least, locally) is obviously close to the crystal structure of $CeAlO_3$ as the $^{27}Al$ satellite line in the $La_{1-x}Ce_xAlO_3$ solid solutions is centered at the same frequency as the $^{27}Al$ spectrum in $CeAlO_3$. These our data confirm that even at low concentration of the $Ce^{3+}$ dopant, it can not be considered as an isolated impurity in $LaAlO_3$. This means that, the 4f ground and 5d exited energy levels of $Ce^{3+}$ substantially depend on the local crystal structure near this ion and, obviously, are located within the valence and conduction bands.

## 4. Conclusions

Detailed $Ce^{3+}$ EPR measurements were done in the $La_{1-x}Ce_xAlO_3$ solid solutions in form of single crystals and powdered samples at Ce concentration from x = 0.001 up to x = 1. The distinct $Ce^{3+}$ EPR signal was observed only at Ce concentration bigger of x = 0.005. This could be related to two facts: (i) main part of the Ce ions was in the non-paramagnetic $Ce^{4+}$ state and (ii) too broad spectral lines due to partial clustering of Ce ions. $Ce^{3+}$ spin Hamiltonian parameters were determined as a function of Ce concentration, including the spin-spin interaction constants between nearest $Ce^{3+}$ ions. It was found that even in the trigonal phase (x = 0.01 –0.1) the $Ce^{3+}$ EPR spectrum is described by g tensor which corresponds to the orthorhombic local symmetry at Ce ion, i.e. incorporation of Ce at La site lowers lattice symmetry in vicinity of the Ce ion. This is in contrast with $Gd^{3+}$ and $Nd^{3+}$ ions which also substitute for La, but their EPR spectrum is described assuming the pure trigonal (axial) symmetry [12,23].

Solid-state NMR measurements (both static and MAS) were done for the $^{27}Al$ and $^{139}La$ nuclei. At low Ce concentration (x = 0.005-0.5), electric quadrupole and chemical shift parameters of these nuclei are in good agreement with those already published for pure $LaAlO_3$ crystal. At higher Ce concentration, spectral lines becomes too broad for precise determination of the quadrupole constant. However, it does not markedly depends on Ce content. In $^{139}$La MAS spectra for Ce concentration x = 0.05 and 0.1, well-resolved NMR peaks from different Ce-O-La bonds are well seen. This fine structure of NMR spectra was fitted by using $180^0$ and $90^0$ Ce-O-La charge transfer passways between $Ce^{3+}$ paramagnetic ions and La ions. It was found that the charge (and spin) transfer over $180^0$ Ce-O-La bonds is approximately two times more effective than that via $90^0$ Ce-O-La bonds in spite of the shorter Ce-La distance for these $90^0$ bonds.

In contrast to $^{139}$La NMR, $^{27}$Al MAS NMR spectrum does not show visible chemical shift related to Fermi contact interaction. However, an additional satellite peak appears in the $^{27}$Al MAS spectra, which intensity linearly increases with increase of Ce concentration in the $La_{1-x}Ce_xAlO_3$ solid solution. The position of this peak does not depends on Ce concentration up to x = 1 suggesting that it originates from regions locally distorted by Ce ions. Crystal structure of these regions is close to the crystal structure of $CeAlO_3$ as the $^{27}$Al satellite line in $La_{1-x}Ce_xAlO_3$ solid solutions is observed at the same frequency as $^{27}$Al NMR line in $CeAlO_3$. This suggests that even at low Ce concentration, Ce ions prefer to create local regions with modified crystal structure that is also reflected in the orthorhombic symmetry of the $Ce^{3+}$ g tensor, which must be axial according to the *R*-3*c* space group of $LaAlO_3$ lattice. Most probably, Ce influences positions of oxygen ions due to completely different outer electronic shells of La ($5d^16s^2$) and Ce ($4f^25d^06s^2$). Note that this effect can not be seen in $^{139}$La spectra due to strong Fermi contact interaction and broad spectral lines.

**Acknowledgments**

The work was supported by the Czech Science Foundation (project No. 20-12885S).

**References**

[1] P.A. Rodnyi, Physical processes in inorganic scintillators, The CRC Press laser and optical science and technology series, CRC Press, Boca Raton, 1997.

[2] M. Nikl and A. Yoshikawa, 2015. Recent R&D Trends in Inorganic Single-Crystal Scintillator Materials for Radiation Detection. Adv. Opt. Mater. 3 (2015) 463-481.

[3] M. J. Weber, J. Appl. Phys. 44 (1973) 3205.

[4] J. Trummer et al., Nucl. Instrum. Methods Phys. Res., Sec. A 551 (2005) 339.

[5] V.M. Goldschmidt, Die Gesetze Der Krystallochemie (The Laws of Cristallochemistry) Naturwissenschaften 14 (1926) 477– 485 (DOI: 10.1007/BF01507527).

[6] X. Zeng, L. Zhang, G. Zhao, J. Xu, Y. Hang, H. Pang, M. Jie, C. Yan and X. He, Crystal growth and optical properties of $LaAlO_3$ and Ce-doped $LaAlO_3$ single crystals. Journal of Crystal Growth 271 (2004) 319.

[7] R.W. Simon, C.E. Platt, A.E. Lee, et al., Low-loss substrate for epitaxial growth of high-temperature superconductor thin films. Appl. Phys. Lett. 53 (1988) 2677.

[8] J. Pejchal, J. Barta, T. Trojek, R. Kucerkova, A. Beitlerova and M. Nikl, Luminescence and scintillation properties of rare-earth-doped LaAlO$_3$ single crystals, Radiation Measurements 121 (2019) 26.

[9] P.W. Peacock and J. Robertson, Journal of Applied Physics 92 (2002) 4712.

[10] P. Dorenbos, IEEE Trans. Nucl. Sci. 57 (2010) 1162.

[11] E. van der Kolk, J.T.M. de Haas, A.J.J. Bos, C.W.E. van Eijk and P. Dorenbos, Luminescence quenching by photoionization and electron transport in a LaAlO$_3$:Ce$^{3+}$ crystal, Journal of Applied Physics 101 (2007) 083703.

[12] W. Low, A. Zusman, Paramagnetic resonance spectrum of gadolinium in LaAlO$_3$, Phys. Rev. 130 (1963) 144-150.

[13] H.J.A. Koopmans, M.M.A. Perik, B. Neiuwenhuijse, P.J. Gellings, The simulation and interpretation of the EPR powder spectra of Gd$^{3+}$ - doped LaAlO$_3$, Phys. Stat. Solidi (b) 122 (1984) 317-330.

[14] H. Luo, A.J.J. Bos, P. Dorenbos, Controlled electron-hole trapping and detrapping process in GdAlO$_3$ by valence band engineering, J. Phys. Chem. C 120 (2016) 5916-5925.

[15] L. Vasylechko, A. Senyshyn, D. Trots, R. Niewa, W. Schnelle, M. Knapp, CeAlO$_3$ and Ce$_{1-x}$R$_x$AlO$_3$ (R = La, Nd) solid solutions: crystal structure, thermal expansion and phase transitions, J. Solid State Chem. 180 (2007) 1277-1290.

[16] M. Buryi, V.V. Laguta, E. Mihokova, P. Novak, M. Nikl, Electron paramagnetic resonance study of the Ce$^{3+}$ pair centers in YAlO$_3$:Ce scintillator crystals, Phys. Rev. B 92 (2015) 224105.

[17] M. Buryi, V. Laguta, M. Nikl, V. Gorbenko, T. Zorenko, Yu. Zorenko, LPE growth and study of the Ce3+ incorporation in LuAlO3:Ce single crystalline Film Scintillators, CrystEngComm 21, 3313-3321 (2019).

[18] J. Pejchal, V. Babin, M. Buryi, et al., Untangling the controversy on Ce$^{3+}$ luminescence in LaAlO$_3$ crystals (submitted to Mater. Chem. C).

[19] S.A. Hayward, F.D. Morrison, S.A.T. Redfern, E.K.H. Salje, J.F. Scott, K.S. Knight, S. Tarantino, A.M. Glaser, V. Shuvaeva, P. Daniel, M. Zhang, M. A. Carpenter, Transformation processes in LaAlO$_3$: neutron diffraction, dielectric, thermal, optical, and Raman studies, Phys. Rev. B 72 (2005) 054110.

[20] J.R. Pilbrow, Transition Ion Electron Paramagnetic Resonance, Clarendon Press, Oxford, 1990.

[21] EPR spectra were simulated using the "Visual EPR" programs by V. Grachev (www.visual-epr.com).

[22] H.R. Asatryan, J. Rosa, J.A. Mares, EPR studies of Er$^{3+}$, Nd$^{3+}$ and Ce$^{3+}$ in YAlO$_3$ single crystals, Solid State Commun. 104 (1997) 5-9.

[23] T. Maekawa, Y. Takahashi, H. M. Shimizu, M. Jinuma, A. Masaike, T. Yabuzaki, A large nuclear polarization of $^{139}$La in Nd$^{3+}$: LaAlO$_3$ for testing the time reversal invariance, Nucl. Instr. and Methods in Phys. Research A 336 (1995) 115-119.

[24] R.J. Birgeneau, M.T. Hutchings, R.N. Rogers, Magnetic interaction between rare-earth ions in insulaters. III. EPR measurements of $Ce^{3+}$ pair-interaction constants in $LaCl_3$, Phys. Rev. 175 (1968) 1116-1133.

[25] P.W. Anderson and P.R. Weiss, Exchange narrowing in paramagnetic resonance, Rev. Mod. Phys. 25 (1953) 269-276.

[26] K.J.D. Mackenzie, M.E. Smith, Multinuclear Solid-State NMR of Inorganic Materials, Pergamon Materials Series, Vol. 6, an Imprint of Elsevier Science, Amsterdam, 2002.

[27] A. Abragam, The principles of nuclear magnetism, Oxford university press, Oxford, 1961.

[28] Yu.O. Zagorodniy, R.O. Kuzian, I.V. Kondakova, M. Marysko, V. Chlan, H. Štěpánková, N.M. Olekhnovich, A.V. Pushkarev, Yu.V. Radyush, I.P. Raevski, B. Zalar, V.V. Laguta, V.A. Stephanovich, Chemical disorder and $^{207}Pb$ hyperfine fields in the magnetoelectric multiferroic $Pb(Fe_{1/2}Sb_{1/2})O_3$ and its solid solution with $Pb(Fe_{1/2}Nb_{1/2})O_3$, Phys. Rev. Mat. 2 (2018) 014402.

[29] NMR spectra were simulated using the Bruker "SOLA" program, www.bruker.com.

[30] K.A. Muller, E. Brun, B. Derighetti, J.E. Drumheller, F. Waldner, Nuclear magnetic resonance of $La^{139}$ and $Al^{27}$ and electron paramagnetic resonance of $Fe^{3+}$ in lanthanum aluminate, Phys. Lett. 9 (1964) 223-224.

[31] L. Bertini, C. Luchinat, G. Parigi, Magnetic susceptibility in paramagnetic NMR, Prog. Nucl. Magn. Reson. Spectrosc. 40 (2002) 249-273.

[32] F. Blanc, D.S. Middlemiss, L. Buannic, J.L. Palumbo, I. Farnan, C.P. Grey, Thermal phase transformations in $LaGaO_3$ and $LaAlO_3$ perovskites: An experimental and computational solid-state NMR study, Solid State Nuclear magnetic resonance 42 (2012) 87-97.

[33] N.C. George, A.J. Pell, G. Dantelle, et al., Local environments of dilute activator ions in the solid-state lighting phosphor $Y_{3-x}Ce_xAl_5O_{12}$, Chemistry of Materials 25 (2013) 3979-3995.